\documentclass[12pt,preprint]{aastex}

\lefthead{Skinner \& G\"{u}del}
\righthead{RW Aur}

\begin{document}

\title{Chandra Resolves the T Tauri Binary System RW Aur}

\author{Stephen L. Skinner\footnote{CASA, Univ. of Colorado,
Boulder, CO, USA 80309-0389; stephen.skinner@colorado.edu} and
Manuel  G\"{u}del\footnote{Dept. of Astrophysics, Univ. of Vienna,
T\"{u}rkenschanzstr. 17,  A-1180 Vienna, Austria; manuel.guedel@univie.ac.at}}

%
\newcommand{\ltsimeq}{\raisebox{-0.6ex}{$\,\stackrel{\raisebox{-.2ex}%
{$\textstyle<$}}{\sim}\,$}}
%
\newcommand{\gtsimeq}{\raisebox{-0.6ex}{$\,\stackrel{\raisebox{-.2ex}%
{$\textstyle>$}}{\sim}\,$}}

\begin{abstract}
RW Aur is a multiple T Tauri system consisting of an early-K type 
primary (A) and a K5 companion (B) at a separation of
1.$''$4.   RW Aur A drives a  
bipolar optical jet that is well-characterized optically. We present
results of a sensitive {\em Chandra} observation whose primary objective
was to search for evidence of soft extended X-ray emission along the 
jet, as has been seen for a few other nearby T Tauri stars. 
The binary is clearly resolved by {\em Chandra} and both stars
are detected as X-ray sources. The X-ray spectra of both stars 
reveal evidence for cool and hot plasma. 
Suprisingly, the X-ray luminosity 
of the less-massive secondary  is at least twice that of  
the primary  and is variable. The  disparity is attributed to
the primary whose X-ray luminosity is at the low end of the
range for classical T Tauri stars of similar mass based on 
established correlations.
Deconvolved  soft-band images show  evidence for slight outward elongation
of the source structure of RW Aur A along the blueshifted jet axis inside
the central arcsecond. In addition, a faint X-ray emission peak is present on the
redshifted axis at an offset of 1.2$''$$\pm$0.$''$2 from the star.
Deprojected jet speeds determined from previous optical studies are too
low to explain  this faint emission peak as shock-heated jet plasma. 
Thus, unless flow speeds in the redshifted jet have been underestimated,
other mechanisms such as magnetic jet heating  may be involved.  
\end{abstract}


\section{Introduction}
RW Aur is a binary system consisting of a pair of 
classical T Tauri stars (TTS). Their properties are 
summarized in Table 1. The optically-bright
$\sim$K1 primary RW Aur A and fainter K5 secondary
were clearly revealed in 
K-band speckle images acquired with the Hale 5m Telescope 
by Ghez, Neugebauer, \& Matthews (1993). A third faint
component identified as C was also visible in their K-band
speckle images at an offset of 0.$''$12 from B but 
component C was not confirmed in subsequent {\em HST} 
optical images which leaves its reality in doubt
(Ghez, White, \& Simon 1997; White \& Ghez 2001). Component B lies at a separation of
1.$''$397 $\pm$ 0.$''$026 from A along P.A. = 254.6$^{\circ}$ $\pm$ 1.0$^{\circ}$  
(White \& Ghez 2001). Low-amplitude periodic variations in optical
spectral features of RW Aur A were reported by Petrov et al. (2001).
They argued that these might be due to a low-mass companion in a
2.77 day orbit but they also noted that the periodic variability
could be due to non-axisymmetric accretion onto a single star.

Hirth et al. (1994) detected an asymmetric bipolar outflow from RW Aur A
using  H$\alpha$ and forbidden-line emission long-slit spectra. 
Follow-up deep imaging in [S II] $\lambda\lambda$6716, 6731
by Mundt \& Eisl\"{o}ffel. (1998) traced the blueshifted southeastern
lobe of the jet-like flow out to more than 100$''$. Even though the
blueshifted lobe is more extended the redshifted lobe was found to
be brighter within 10$''$ of the star. Their images revealed at least
ten knots in the jet. The bright inner redshifted lobe  was confirmed in
spectacular [S II] and [O I] adaptive optics images obtained in  1997 with the CFHT
by Dougados et al. (2000) and  in  1998 by L\'{o}pez-Mart\'{i}n, Cabrit, 
\& Dougados (2003).  Their images traced
the redshifted lobe down to separations of $<$1$''$ and showed several  
jet knots including a prominent redshifted knot designated R5 at a 
separation of 1.$''$07 - 1.$''$32 from the star. They determined the 
position angle of the inner regions of the blueshifted jet to be 
P.A. = 130$^{\circ}$ $\pm$ 2$^{\circ}$.
{\em HST} STIS observations were obtained by Woitas et al. (2002)
in which the jet was traced down to within 0.$''$1 of the 
central source in forbidden emission lines.
 Their results confirm the radial velocity asymmetry with the outflow
velocity of the blueshifted lobe being about 50\% larger than
the redshifted lobe and suggest that the asymmetry difference
is time-variable on timescales of a few years. Further analysis
of the {\em HST} data by Melnikov et al. (2009) shows evidence
for jet knots at separations ranging from  $<$1$''$ out to $\approx$3$''$
from the star and provides estimates of  jet properties (e.g. electron 
temperature and density, ionization fraction, radial velocity,
and jet mass-loss rate).  

RW Aur and its jet have been well-studied optically but the
availability of modern X-ray telescopes with arcsecond spatial
resolution such as {\em Chandra} has now opened a new spectral window
for jet studies. Jets with outflow speeds of several hundred km s$^{-1}$
can produce shocks with temperatures of a few MK capable of 
emitting soft X-rays. In addition, other processes
such as entrained magnetic fields may play a role in  heating 
young stellar jets to X-ray temperatures.

Observational evdidence for X-ray emitting jets in young stars
has emerged over the past decade, but so far only a few examples
are known. The best-studied example to date is the classical TTS
DG Tau in  which faint soft  X-ray emission (E $<$ 2 keV) extending outward 
$\approx$4$''$ - 5$''$ from the star along the jet was seen
in high-sensitivity  {\em Chandra} images (G\"{u}del et al. 2005; 2008).
Other examples include the optically-revealed stars 
RY Tau (Skinner, Audard, \& G\"{u}del 2011), Z CMa (Stelzer et al. 2009), and 
HD 163296 (Swartz et al. 2005; G\"{u}nther, Schneider, \& Li 2013)
as well as heavily-obscured protostars such as L1551 IRS 5 (Bally,
Feigelson, \& Reipurth 2003; Schneider, G\"{u}nther, \& Schmitt 2011). 
The X-ray jet emission from these young stars and protostars is typically 
faint (tens of counts in exposures of
$<$100 ksec), soft (E $<$ 2 keV), and visible only out to a 
few arcseconds from the central star.

In order to better characterize the ubiquity and nature of X-ray jet 
emission from young stars, additional examples of this phenomenon
are needed.  RW Aur offers a promising candidate for X-ray jet
studies because of its proximity (d = 140 pc), low intervening
extinction (A$_{v}$ = 0.4 mag; Table 1), and well-delineated
optical jet which has been traced inward to within $\approx$1$''$
of the star. A previous {\em XMM-Newton} observation in 2007 detected 
RW Aur as an X-ray source (G\"{u}del et al. 2010) but the angular resolution 
was not sufficient to clearly resolve the close AB pair.
We present here a higher spatial resolution  {\em Chandra}
observation of RW Aur which clearly resolves the binary
and shows that both stars are X-ray sources.
This observation reveals
previously unknown X-ray properties of RW Aur A including a faint 
emission peak lying on the redshifted jet axis near the 
star that may be jet-related as well as an unusual X-ray
spectrum that includes very soft and very hard emission.
An unexpected finding   is that the less-massive
secondary RW Aur B is more luminous  in X-rays
than the jet-driving primary.

\section{Chandra Observation}

The {\em Chandra} observation (ObsId 14539) was carried out on
12 January 2013 from 01:13 - 18:09 TT with an exposure live time
of 54,482 s.  Exposures were obtained using the ACIS-S (Advanced CCD
Imaging Spectrometer) array in FAINT  timed-event mode.
RW Aur was placed at the nominal aimpoint on the  ACIS-S3 CCD which
was configured  in 1/8 subarray mode. The use of subarray mode
restricts the field-of-view to a rectangular region of
128 $\times$ 1024 native pixels, or $\approx$63$''$ $\times$ 504$''$
at the native pixel size of 0.$''$492. Subarray mode was selected
in order to use a short 0.4 s frame time that would mitigate 
photon pileup in the event of a large X-ray flare (no large
flares occurred). For an on-axis point source,
the ACIS-S 90\%  encircled energy radius at 1.4 keV  is
R$_{90}$ $\approx$ 0.$''$9.
Further information on {\em Chandra} and its instrumentation can
be found in the {\em Chandra} Proposer's
Observatory Guide (POG)\footnote {See http://asc.harvard.edu/proposer/POG}.

The pipeline-processed data  files provided by the {\em Chandra} X-ray
Center (CXC) were  analyzed using standard science
threads with CIAO version 4.4\footnote{Further information on
{\em Chandra} Interactive
Analysis of Observations (CIAO) software can be found at
http://asc.harvard.edu/ciao.}.
The CIAO processing  used recent calibration
data from CALDB version 4.4.10.
Source events, spectra, and light curves were extracted from circular regions of
radii 0.$''$9 centered on the X-ray peaks of A and B. Because of the close
1.$''$4 source separation, the common region between A and B where the 
two extraction circles overlap was excluded in order to minimize
cross-contamination. 
Background was extracted from nearby source-free  regions.
Background is negligible,
amounting to less than 1 count (0.2 - 8 keV) within the extraction
circle during the 54 ks exposure.
CIAO {\em specextract} was used to extract
spectra along with source-specific
response matrix files (RMFs) and auxiliary response files (ARFs).
Spectral fitting, timing analysis, and image analysis were undertaken with the HEASOFT
{\em Xanadu}\footnote{http://heasarc.gsfc.nasa.gov/docs/xanadu/xanadu.html.}
software package (v. 6.12) including XSPEC vers. 12.7.1 and  XRONOS vers. 5.2.1.
Additional tests  for source variabilility
were carried out on  energy-filtered source event lists using
the Bayesian-method CIAO tool {\em glvary}
(Gregory \& Loredo 1992, 1996).

\section{Results}

Both RW Aur A and B are clearly detected. 
Their X-ray properties are summarized in Table 2.

\subsection{Image Analysis}

A key motivation for this observation was to determine whether
the RW Aur A jet  (in addition to the star itself) might be a source of
X-rays. We have thus carefully examined the {\em Chandra} images
to discern whether any extension is visible outward along the
direction of the optical jet. Based on {\em Chandra} observations
of other jet sources such as DG Tau (G\"{u}del et al. 2005; 2008)
and RY Tau (Skinner, Audard, \& G\"{u}del 2011) we anticipate
that any extended X-ray structure from the jet would be dominated 
by soft emission with   energies E $<$ 2 keV and would 
be visible out to  a few arcseconds from the central source.
The redshifted axis along P.A. = 310$^{\circ}$ is of particular
interest since optical images show at least six redshifted emission 
knots at separations of 0.$''$25 - 3.$''$5 from the star 
(Dougados et al. 2000; Melnikov et al. 2009).

{\em Chandra} clearly separates the A and B components as evident
from the soft-band (0.2 - 2 keV) image in Figure 1 (without deconvolution).
If any significant X-ray extension were present
we would expect to see source elongation at offsets of $>$1$''$ from the 
central peak where PSF effects no longer dominate. No  such structure
is visible but a slight outward elongation of the contours 
along the blueshifted jet is seen inside the central arcsecond.
We find no significant difference between the X-ray centroid
position of RW Aur A in the soft (0.2 - 2 keV) and hard (2 - 8 keV) band
images.

To search more closely for evidence of extended structure we generated
deconvolved soft-band images using the CIAO 
$arestore$\footnote{http://cxc.harvard.edu/ciao/ahelp/arestore.html}
tool which implements
the Lucy-Richardson method (Lucy 1974; Richardson 1972).
Deconvolution removes some of the blurring effect due to telescope optics.
However, the procedure is not perfect due to several factors including 
imprecise modeling  of the {\em Chandra} telescope optics and PSF as well
as uncertainties in the energy and off-axis dependence of the PSF.
In particular, there is a suspected asymmetry in the {\em Chandra} PSF
which is not yet taken into account in {\em Chandra} PSF simulation tools,
as discussed further below. It is also worth noting that the  
Lucy-Richardson method gives best results for bright point sources
whereas the number of counts in the present case is modest, 
especially for source A (Fig. 1). 
The deconvolution was performed using an observation-specific image of the 
{\em Chandra} PSF produced using the {\em Chart} ray-tracing and {\em MARX} simulators
as prescribed in CIAO science thread procedures. These procedures take
into account the  telescope optics as well as the position of the source on the detector 
and the  source spectrum. Event lists provided by CXC now include energy-dependent subpixel
event repositioning (EDSER) by default. Since {\em Chandra} simulation tools do not
yet have the capability to generate  EDSER-enhanced PSFs, we removed the EDSER
prior to generating simulated PSFs.
 
Since structure in deconvolved images can be sensitive to the 
number of iterations applied during the procedure, we generated 
images using 25, 50, 100, and 200 iterations.
Representative deconvolved images in the 0.2 - 2 keV and
0.2 - 1.5 keV energy bands  using 100 iterations and
the respective PSF images used to produce them are shown in Figure 2. 
The structure of RW Aur A is now almost completely confined
to within a region of radius 0.$''$5 due to mitigation of PSF smearing 
effects (Fig. 2-top left).
Along the blueshifted jet, some  weak extension of the central core region
out to 0.$''$6 from the peak of source A is visible in the deconvolved
images, as was already  noticeable in the raw image without 
deconvolution  (Fig. 1-left). The blueward extension becomes more
apparent in the smoothed deconvolved image using the more restrictive
0.2 - 1.5 keV energy cut shown in Figure 2-bottom.
No corresponding extension outward from the central source is visible along
the redshifted jet, but a faint isolated emission peak is present along
the redshifted axis at an offset of 1.$''$2$\pm$0.2$''$ from
the central peak.  This weak feature is also present in images
obtained using 50 and 200 iterations and becomes more localized
as the number of iterations is  increased from 50 to 200.
The feature contains 11$\pm$2 counts (0.2 - 2 keV), where the
uncertainty reflects only the dependence on the number of iterations
and is not statistical.

Some caution is needed in interpreting the faint X-ray
peak on the redshifted jet axis as a real emission feature.
There is a known asymmetry in the  {\em Chandra} 
PSF\footnote{Additional information on the
PSF asymmetry  can be found at:~ \\
http://cxc.harvard.edu/ciao/caveats/psf\_artifact.html .}
which can produce artificial structure in
deconvolved images inside the central arcsecond.  
The  artificial structure is hook-shaped, contains a small fraction
of the total flux, and is confined to small offsets  of 
0.$''$6 $\leq$ $r$ $\leq$ 1$''$ from the central source
over a range of roll-dependent position angles. The asymmetry does
not affect images on scales larger than one arcsecond.
For the RW Aur observation the nominal roll angle was
ROLL = 272.1$^{\circ}$ and the affected region lies along  
P.A. = 282.9$^{\circ}$ ($\pm$25$^{\circ}$), measured from north to east.
The sector regions that may be affected by the PSF 
asymmetry\footnote{The sector regions were generated using the CIAO 
tool $make\_psf\_asymmetry\_region$ .} 
are overplotted in Figure 2 for both the A and B components.
We refer to these hereafter as  sector A and sector B.

For deconvolved images obtained using $\geq$50 iterations,
no significant counts  are present in sector A
but residual counts are clearly present in the sector B.
In the deconvolved image in Figure 2-top left (100 iterations; 0.2 - 2 keV),
there is $<$1 count in sector A and $\approx$41 counts
in sector B. The elongated structure in B does not
extend beyond the outer sector boundary at $r$ = 1$''$
and the 41 counts in sector B lie within the lower
P.A. range of the sector. On the other hand, the
faint feature visible in A lies beyond the 
outer sector boundary and appears on the redshifted
jet axis at the higher P.A. end of  sector A.
The above differences in the  spatial distribution 
of counts between sources A and B combined with 
the fact that sector A is free of counts suggests
that the weak feature at $r$ $\approx$ 1.$''$2 along the
redshifted jet axis of RW Aur A could be real
X-ray emission associated with the jet. In order
to confirm this, a follow-up observation obtained
at a different roll angle in order to rotate
the region affected by the asymmetry away from the 
redshifted jet axis would be needed.

To make a brief comparison of the deconvolved X-ray images with
previous optical observations, we note that the 1.$''$2 offset 
of the X-ray feature from source A is nearly identical to that of
the redshifted knot R5 seen in December 1998 CFHT
observations (Dougados et al. 2000; L\'{o}pez-Martin et al. 2003).
However,  they find a proper motion of $\mu$ = 0.$''$24$\pm$0.$''$05 yr$^{-1}$
for the R5 knot so the X-ray feature seen in our January 2013
observation cannot be  R5. A similar argument applies to redshifted
knots J1 (offset 0.$''$25), J2 (0.$''$9), and J3 (1.$''$8)
seen in December 2000 {\em HST} images (Melnikov et al. 2009)
whose proper motions were estimated to be $\approx$0.$''$2 yr$^{-1}$
by Woitas et al. (2002). Based on these proper motions, the above
knots would have moved outward $\approx$2.$''$4 during the
$\approx$12 years between the {\em HST} and {\em Chandra}
observations. If the faint X-ray peak originates in the jet
then it is a new feature that has formed since the previous
CFHT and {\em HST} observations.

\subsection{Timing Analysis}

Light curves of RW Aur A  and B are shown in Figure 3.  
The CIAO $glvary$ tool gives a probablility of constant count rate
P$_{const}$ = 0.96 (0.2 - 8 keV) for RW Aur A, so no significant
variability was detected in the primary star.
On the other hand, a similar test for RW Aur B gives 
P$_{const}$ $<$ 0.001 (0.2 - 8 keV) so variability is present.
As the light curves in Figure 2 show, the broad-band count rate of 
B increased by $\approx$30\%  in the second half of the observation.
This increase is most clearly seen in the hard 2 - 8 keV band.
A slight increase in the soft band (0.2 - 2 keV) count rate is 
also seen, but no variability was detected in the softest emission
in the 0.2 - 1 keV range.
The increase in count rate of component B was accompanied by a 
flux increase (Sec. 3.3.2) and a hardness increase. During the 
first half of the observation 
the median photon event energy of B was E$_{50}$ = 1.07 keV
and during the second half it was E$_{50}$ = 1.17 keV.
The corresponding mean photon energies were 
$\overline{\rm E}$ = 1.29 keV and 1.45 keV.

\subsection{Spectral  Analysis}

The spectra of RW Aur A and B are shown in Figure 4.
Spectral fits using absorbed optically-thin plasma models
require two temperature components (2T) to obtain acceptable
fits as measured by $\chi^2$ statistics. In the case of
RW Aur A we also consider a hybrid model consisting of 
a cool isothermal plasma (1T) plus a power-law to
reproduce the emission at higher energies. 
Fit results are given in Table 3. 

\subsubsection{RW Aur A}

The best-fit X-ray absorptions toward both A and B are quite
low. In the case of RW Aur A the absorption is so low that
the  value of the column density N$_{\rm H}$ cannot be
accurately determined from the X-ray spectrum because of the
rapid falloff in  ACIS-S sensitivity  at energies
below 0.5 keV. Based on spectral fits we derive
an upper 90\%  confidence bound 
N$_{\rm H}$ = 6.5 $\times$ 10$^{20}$ cm$^{-2}$. 
Table 3 includes a fit in which N$_{\rm H}$ was allowed to
vary as well as fits in which it was held fixed at 
the  upper 90\%  confidence bound.
The above upper limit corresponds
to A$_{\rm V}$ = 0.3 mag using the N$_{\rm H}$ to
A$_{\rm V}$ conversion of Gorenstein (1975) or
A$_{\rm V}$ = 0.4 mag using the conversion of
Vuong et al.  (2003). These values are consistent
with the extinction obtained in ground-based
observations by White \& Ghez (2001), namely
A$_{v}$ = 0.39$\pm$0.33 or  equivalently
A$_{\rm V}$ = 0.35$\pm$0.30 mag.  

The spectrum of RW Aur A has some noteworthy features.
High-temperature lines are visible
including the blended He-like triplets Mg XI (1.35 keV; log T$_{max}$ = 6.8),
Si XIII (1.86 keV; log T$_{max}$ = 7.0), and
S XV (2.46 keV; log T$_{max}$ = 7.2) where T$_{max}$ (K) is the
maximum line-power temperature. Thus, there is little doubt
that hot plasma (T $\gtsimeq$ 10 MK) is present in the spectrum. 
Even so,  spectral fits with multi-temperature thermal models
do not tightly constrain the temperature  of the hot plasma. 
Our models are only able to constrain the 90\% confidence lower
bound which is kT$_{2}$ $\geq$ 20 keV if the spectra are rebinned
to a minimum of 10 counts per bin (Table 3) or
kT$_{2}$ $\geq$ 16 keV if rebinned to 20 counts per bin.
The high temperature component is present even in spectra
extracted using circular regions as small as $r$ = 0.$''$3 
centered on RW Aur A. Of the 986 broad-band counts (0.2 - 8 keV)
detected in RW Aur A, 50 (5.1\%) are in the high-energy 4 - 8 keV
range and we estimate no more than 3 of these 50 are due to
contamination from RW Aur B. By comparison, RW Aur B has
2909 broad-band counts of which 70 (2.4\%) are in the
high-energy 4 - 8 keV range.
The very high temperature of RW Aur A is  somewhat unusual since hot-component 
plasma temperatures in T Tauri stars in the absence  of large
flares are typically less than $\sim$5 keV. The inability
to tightly constrain the hot-component temperature is due
at least in part to the limited number of counts in the 
very hard 4 - 8 keV band.

Because of the apparent very high plasma temperature inferred
from 2T thermal models we also tried to fit the spectrum of
RW Aur A with a composite  model consisting of a cool
thermal plasma plus a power-law (PL) to reproduce  the hard continuum.
The physical relevance of a PL model is questionable since TTS
X-ray spectra are usually well-fitted with thermal models.
Nevertheless, the hybrid 1T$+$PL  model provides a slightly better fit
to the spectrum as measured by $\chi^2$ statistics than does
the 2T thermal model (Table 3). In this composite model
the thermal component accounts for almost all of the
observed emission below 2.5 keV and the power-law 
reproduces the hard featureless emission at higher energies (Fig. 5). 
The power-law spectrum is nearly flat  with a photon
power-law index $\Gamma_{ph}$ = $-$0.04 [$-$0.76 - $+$0.59].

The 2T and 1T $+$PL models give similar X-ray luminosities
of log L$_{x}$ = 29.44 - 29.51 (ergs s$^{-1}$) for RW Aur A.
We compare these values with predictions based on 
known correlations between L$_{x}$ and stellar 
mass and  luminosity in Section 4.2.

\subsubsection{RW Aur B}

There are sufficient counts in the {\em Chandra}
spectrum of RW Aur B  to reliably determine a best-fit value of 
N$_{\rm H}$ (as opposed to only an upper limit).
The 2T model in Table 3 in which N$_{\rm H}$
was allowed to vary gives N$_{\rm H}$ =
4.3 [2.3 - 6.5] $\times$ 10$^{20}$ cm$^{-2}$
where brackets enclose the 90\% confidence range.
This equates to A$_{\rm V}$ = 0.20 [0.10 - 0.30] 
(Gorenstein et al. 1975) or 
A$_{\rm V}$ = 0.27 [0.14 - 0.41] (Vuong et al.  2003).
These values suggest low absorption similar to that of
RW Aur A and agree with the value
A$_{\rm V}$ = 0.32$\pm$0.11  for RW Aur B given
in Ghez et al. (1997). However, this value is
significantly less than
A$_{v}$ = 1.56$\pm$0.24 (A$_{\rm V}$ = 1.40$\pm$0.22)
obtained by White \& Ghez (2001). 
For completeness, we have thus included a fit of RW Aur B in Table 3
holding the absorption  fixed at  
N$_{\rm H}$ = 2.24 $\times$ 10$^{21}$ cm$^{-2}$
corresponding to A$_{\rm V}$ = 1.40 using
the Vuong et al. (2003) conversion. This higher
absorption fit yields slightly lower plasma temperatures
and higher L$_{x}$. The reduced $\chi^2$ value is
greater than for the low-absorption fit but is
still marginally acceptable.

The X-ray spectrum of RW Aur B is quite typical of
a cTTS. Hot plasma at kT$_{2}$ $\approx$ 2 - 3 keV is present 
as revealed by  Mg XI, Si XIII, and S XV features in the spectrum (Fig. 3-top).
The  count rate  increased during the second half
of the observation and the emission became slightly
harder as gauged by a change in the  hardness ratio (H.R.; Table 2) 
from H.R. = $-$0.79 in the first half to  $-$0.68 in the second half.

To search for changes in the spectrum of RW Aur B, we extracted and
fitted separate spectra for the first  half (1246 counts) and second 
half (1663 counts) of the observation. 
The spectra were fitted with an absorbed 2T thermal plasma model
and the absorption was held fixed  at the best-fit
value N$_{\rm H}$ = 4.3  $\times$ 10$^{20}$ cm$^{-2}$ 
determined from  the spectral fit for 
the full exposure (Table 3). Fixing the value of
N$_{\rm H}$ is justified since any significant change
in N$_{\rm H}$ would have resulted in different count rates
at very low energies $<$0.5 keV between the first and
second halves, but no significant change was observed.

Fits of the spectra for the first and second halves show
no significant change in the temperature of the 
cool component (kT$_{1}$). In contrast, a modest increase in the temperature 
of the hot plasma component from kT$_{2}$ = 2.87 [2.04 - 6.38, 90\% conf.] keV
to kT$_{2}$ = 3.40 [2.72 - 5.33] keV is inferred. 
The broad-band flux (0.2 - 8 keV)
increased by $\approx$43\% during the second half of
the observation (Table 3 notes). The unabsorbed
flux of RW Aur B during the first half  
was a factor of $\sim$2 greater than that of RW Aur A
and  a factor of $\sim$3 greater in the second half.
Assuming A and B lie at the same distance this implies
that the  secondary is more luminous in X-rays
than the primary.

\vspace*{0.5in}

\section{Discussion}

\subsection{X-ray Jet Emission?}

In general, the conditions for detecting X-ray jet emission are more
favorable in the approaching blueshifted jet because of
lower line-of-sight absorption, but faint X-ray emission in 
both the  redshifted and blueshifted jets 
of the TTS DG Tau has been detected (G\"{u}del et al. 2005; 2008).
Also, RW Aur is somewhat unusual in that its  redshifted jet
is optically much brighter in regions close to the star than the
blueshifted jet (Fig. 1 of Dougados et al. 2000).
Given the above, it is worthwhile to consider whether the
faint X-ray peak along the redshifted jet axis
at an offset of $\approx$1.$''$2 from RW Aur A might
originate in the jet.
 
One possibility is that the faint X-ray feature is 
shock-heated plasma in the jet. The predicted temperature
for a shock-heated jet with a shock speed $v_{s}$ is
(Raga et al. 2002)~T$_{s}$ = 0.15 ($v_{s}$/100 km~s$^{-1}$)$^{2}$ MK.
The spatially-resolved forbidden emission line  data from the {\em HST} 
STIS observations (Melnikov  et al. 2009) give a radial velocity
for the redshifted jet at an offset of $\approx$1.$''$2 from the 
star of $v_{r}$ $\approx$ 100 km s$^{-1}$. In fact they find
little change  in the radial velocity out to offsets of $\approx$3$''$. 
Using the jet inclination angle relative to the line-of-sight
of $i$ = 46$^{\circ}$$\pm$3$^{\circ}$ (L\'{o}pez-Martin et al. 2003)
the deprojected jet velocity is $v_{jet}$ $\approx$ 144$\pm$8 km s$^{-1}$.
Adopting the upper limit of this range and assuming the 
jet impacts a stationary target
($v_{s}$ $\approx$ $v_{jet}$ $\approx$ 152 km s$^{-1}$),
then the maximum shock temperature
is $T_{s}$ $\approx$ 0.35 MK (k$T_{s}$ $\approx$ 0.03 keV).
This temperature is at least a factor of $\sim$4 lower than
that needed to produce detectable thermal X-ray emission
at  kT $\approx$ 0.1 - 0.2 keV (T $\approx$ 1 - 2 MK),
below which {\em Chandra} has little sensitivity.

As an additional check, the predicted intrinsic 
X-ray luminosity of a bow shock is (Raga et al. 2002)

\begin{equation}
L_{jet} = C_{o}\left[\frac{n_{o}}{100~ \rm{cm^{-3}}}\right]^{\alpha}\left[\frac{r_{bs}}{10^{16}~ \rm{cm}}\right]^{\beta}\left[\frac{v_{s}}{100~ \rm{km~ s^{-1}}}\right]^{\gamma} L_{\odot}
\end{equation}

where $n_{o}$ is the preshock number density,
$r_{bs}$ is the characteristic radius of the bow shock around
its axis,  and $v_{s}$ is the shock speed. For the case of a
radiative shock: 
$C_{o}$ = 4.1 $\times$ 10$^{-6}$, $\alpha$ = 1, $\beta$ = 2, $\gamma$ = 5.5.
For the nonradiative case:
$C_{o}$ = 4.5 $\times$ 10$^{-5}$, $\alpha$ = 2, $\beta$ = 3, $\gamma$ = 1. 
Optical observations of  the redshifted jet of RW Aur at an offset of 
1.$''$2 from the star give
$n_{o}$ $\approx$ 1585 cm$^{-3}$ (Melnikov et al. 2009) and 
FWHM $\approx$ 28 AU (Dougados et al. 2000; Woitas et al. 2002).
Taking $r_{bs}$ = 14 AU (i.e. the jet half-width) and assuming
$v_{s}$ $\approx$ 150 km s$^{-1}$ as above, the radiative
case gives $L_{jet,r}$ = 1 $\times$ 10$^{26}$ ergs s$^{-1}$
and for the nonradiative case $L_{jet,nr}$ = 6 $\times$ 10$^{24}$ ergs s$^{-1}$. 
Assuming that the weak $\approx$11 count X-ray peak is a thermal source
seen through negligible absorption its intrinsic X-ray luminosity is
L$_{x}$ =  3 $\times$ 10$^{27}$ ergs s$^{-1}$ assuming 
kT = 0.2 keV or L$_{x}$ =  2 $\times$ 10$^{27}$ ergs s$^{-1}$ assuming
kT = 0.5 keV. These L$_{x}$ values are at least an order of magnitude
greater than predicted above for a shocked jet.

The above comparisons  show that the faint X-ray peak is
not compatible with emission from a shocked jet knot
{\em unless} the  redshifted jet (shock) velocity is significantly greater than 
determined from previous optical data. In order to reconcile the
X-ray emission with a shock interpretation one would need to postulate the
existence of optically-undetected jet material moving at 
speeds in excess of $v_{jet}$ $\sim$ 300 km s$^{-1}$ or a jet inclination
angle $i$ $>$ 70$^{\circ}$. Even though deprojected speeds in the
blueshifted jet may approach $-$275 km s$^{-1}$  assuming  the nominal
jet inclination $i$ = 46$^{\circ}$$\pm$3$^{\circ}$    (Melnikov et al. 2009), 
there is presently no firm evidence for such high speeds in the redshifted jet or
for a jet inclination $>$ 70$^{\circ}$. Thus, based on existing data,
a shock origin for the faint X-ray peak seems unlikely and 
other explanations must be considered.  

If the jet is threaded by a weak magnetic field such as
detected in the HH 80-81 jet (Carrasco-Gonz\'{a}lez et al. 2010),
then shock-heating could be augmented by magnetic heating to
achieve X-ray temperatures. Another possible mechanism for attaining
higher outflow speeds and X-ray emitting temperatures  
is via plasmoid ejections (Hayashi et al. 1996).
Using published mass and radius estimates for RW Aur A
(Table 1), the escape speed from its surface is 
$v_{esc}$ $\approx$ 560 km s$^{-1}$
so any ejected plasmoid would be moving
at speeds sufficient to produce soft X-ray
emission via shocks or even intrinsic 
thermal X-ray emission if the plasmoid
is hot enough (T $\gtsimeq$ 10$^{7}$ K).
Further discussion of the plasmoid hypothesis
can be found in Skinner et al. (2011).

\subsection{X-ray Luminosity Disparity}

As noted above, the less-massive component RW Aur B
is $\sim$2 - 3 times more X-ray luminous than RW Aur A.
This is contrary to expectations since L$_{x}$
is correlated with stellar mass M$_{*}$ and 
stellar luminosity L$_{*}$ in TTS. The reason
for the correlation is not well-understood but it
has been documented in large nearly-coeval samples of TTS
in Taurus (Telleschi et al. 2007) and in the Orion 
Nebula Cluster (Preibisch et al. 2005). 
For the Taurus cTTS sample studied by Telleschi et al., 
which is relevant here, their regression fit using
the parametric estimation maximization method gives 
log L$_{x}$ = (1.70$\pm$0.20)log(M$_{*}$/M$_{\odot}$) 
$+$ 30.13$\pm$0.09. For the stellar luminosity correlation
they obtain
log L$_{x}$ = (1.16$\pm$0.09)log(L$_{*}$/L$_{\odot}$)
$+$ 29.83$\pm$0.06.

Adopting M$_{*}$ = 1.34$\pm$0.18 M$_{\odot}$ for RW Aur A
(White \& Ghez 2001), the above relation predicts 
log L$_{x}$ = 30.35 (30.14 - 30.66), where the range in
parentheses here and below takes into account both the uncertainties 
in the regression fit and M$_{*}$.  The lower
value M$_{*}$ = 0.9 M$_{\odot}$ from Ingleby et al. (2013)
gives log L$_{x}$ = 30.05 (29.95 - 30.25).
Both of the above estimates are higher than the observed
value log L$_{x}$ = 29.4 - 29.5 but the estimate for the 
smaller mass M$_{*}$ = 0.9 M$_{\odot}$ is in better agreement
with the X-ray data. Thus, either the X-ray luminosity of RW Aur A
is at the low end of the expected range for cTTS of 
similar mass (the scatter is large; see Fig. 1 of Telleschi et al. 2007)
or its mass has been slightly overestimated. 
A similar calculation based on L$_{*}$ yields  good agreement
with the observed L$_{x}$ for the lower stellar luminosity
L$_{*}$ = 0.5 L$_{\odot}$ given by Ingleby et al. (2013).
Likewise, reasonably good agreement is obtained with the
correlation predictions for RW Aur B if the lower limits
on its M$_{*}$ and L$_{*}$ are adopted (Table 1; White \& Ghez 2001).

\subsection{Accretion Shocks}

Material accreting onto the star can produce soft X-rays
in shock-heated plasma at temperatures of a few MK. Soft emission 
below 0.5 keV is present in the spectrum of RW Aur A (Fig. 4 bottom)
with 22 counts detected with energies E $<$ 0.5 keV. 
Since RW Aur A is accreting at a rate of
$\dot{\rm M}_{acc}$ $\approx$ 2 $\times$ 10$^{-8}$ M$_{\odot}$ yr$^{-1}$
(Ingleby et al. 2013), we provide a  brief comparison
of accretion shock estimates with the observed X-ray
properties.

The predicted temperature of the post-shock plasma is (Ingleby et al. 2013)
\begin{equation}                                                                                                                             \
T_{s,acc} = 8.6 \times 10^{5}\left[\frac{M_{*}}{0.5 M_{\odot}}\right]\left[\frac{R_{*}}{2 R_{\odot}}\right]^{-1} ~K.                                               \end{equation}
Taking M$_{*}$ = 0.9 M$_{\odot}$ and R$_{*}$ = 1.1 R$_{\odot}$ (Table 1) the above
gives T$_{s,acc}$ = 2.8 MK or kT = 0.24 keV.  The intrinsic (unabsorbed) emission of
a thermal  plasma at this temperature occurs mostly at E $<$ 1 keV with the 
flux peaking near 0.65 keV. Thus,  some of the soft
emission detected in RW Aur A may be accretion-related. 

We have simulated an accretion shock contribution  using a 3T $vapec$
model. This model  is identical to  the 2T model in
Table 3 with N$_{\rm H}$ held fixed except that an additional very  
cool component at a fixed temperature
kT$_{acc}$ = 0.24 keV has been added to simulate an accretion shock.
The emission measure ($norm$) of this component was allowed to vary.
The 3T fit converges to values very similar to those of the 2T model
and the accretion component contribution  is small.
Specifically, F${_x,acc}$(0.2 - 8 keV) = 
1.42 (2.35) $\times$ 10$^{-15}$ ergs cm$^{-2}$ s$^{-1}$ where the
unabsorbed value in parentheses gives log L$_{x,acc}$ = 27.74 ergs s$^{-1}$.
The simulated accretion component contributes
only $\approx$15 soft counts (E $\leq$ 1 keV) to the observed 
spectrum and its associated L$_{x}$ is much less than the 
total accretion luminosity 
L$_{acc}$ $\sim$ (GM$_{*}$$\dot{\rm M}_{acc}$)/R$_{*}$ $\sim$
10$^{33}$ ergs s$^{-1}$.

\section{Summary}

The main results of this study are the following:

\begin{enumerate}

\item {\em Chandra} has detected X-ray emission from 
RW Aur A and its companion RW Aur B. The companion
is more luminous in X-rays and its emission is variable.
Both stars show cool and hot plasma in their spectra.
In order to reproduce the hot component of RW Aur A,
either a very high temperature thermal plasma or
a power-law continuum is required. An accretion
shock could be responsible for a small fraction of the 
coolest emission in RW Aur A at E $<$ 1 keV.

\item Deconvolved soft-band images show slight X-ray extension  
in the source structure of RW Aur A along the blueshifted
jet axis close to the star. In addition, deconvolved images 
reveal a faint X-ray emission peak lying 1.$''$2 $\pm$ 0.$''$2 from 
the star along the  redshifted jet axis.  
This X-ray peak is located at an offset where an 
optical emission  knot was previously observed and it lies 
outside the region that can be affected by artificial PSF-related 
structure. As such, the X-ray peak may be real emission associated
with the redshifted jet but since the emission is faint a 
{\em Chandra}  observation acquired at a different
roll angle is needed for confirmation.
A comparison  with shock models shows that this peak is unlikely 
to be due to shock-heated jet plasma unless the redshifted jet speed 
is substantially higher than determined
from previous optical observations. Other mechanisms may thus be at work
such as magnetic jet heating  or plasmoid ejections.

\item  The X-ray luminosity of RW Aur A is significantly
less than  predicted  based on its published mass range (Table 1) 
and a known correlation L$_{x}$ $\propto$ M$_{*}$ in  TTS.
This suggests that L$_{x}$ for RW Aur A is at the low end of
the range for cTTS of similar mass (the scatter for a given 
mass is large) or that its mass has been slightly overestimated.
The predicted L$_{x}$ for RW Aur A based on a correlation with
stellar luminosity agrees well with the observed L$_{x}$ if
L$_{*}$ $\approx$ 0.5 L$_{\odot}$. For RW Aur B, satisfactory
agreement between the observed L$_{x}$ and correlation 
predictions is obtained.

\end{enumerate}  

\acknowledgments

This work was supported by {\em Chandra} award GO3-14007X
issues by the Chandra X-ray Observatory Center (CXC). The CXC is operated by the
Smithsonian Astrophysical Observatory (SAO) for, and on behalf of,
the National Aeronautics Space Administration under contract NAS8-03060.

\newpage

                                                                                                                       
\begin{deluxetable}{lccccccccc}
\tabletypesize{\scriptsize}
\tablewidth{0pt}
\tablecaption{Properties of RW Aur}
\tablehead{
           \colhead{Object}             &
           \colhead{Type}               &
           \colhead{log Age}                &
           \colhead{M$_{*}$}            &
           \colhead{R$_{*}$}            &
           \colhead{T$_{eff}$}          &
           \colhead{log L$_{*}$}            &
           \colhead{A$_{v}$}            &
           \colhead{d}                  &
           \colhead{Refs.}             \\
           \colhead{}                   &
           \colhead{}                   &
           \colhead{(y)}                   &
           \colhead{(M$_{\odot}$)}         &
           \colhead{(R$_{\odot}$)}         &
           \colhead{(K)}                   &
           \colhead{(L$_{\odot}$)}                   &
           \colhead{(mag)}                   &
           \colhead{(pc)}                &
           \colhead{}
}
\startdata
RW Aur A &  K3         & ...            &  0.9            & 1.1    & ...        & $-$0.30             & 0.55          \
   & 140 & 1,2    \\
RW Aur A &  K1$\pm$2   & 6.92$\pm$0.31  &  1.34$\pm$0.18  & ...   & 5082       & 0.23$\pm$0.14       & 0.39$\pm$0.33  \
  & 140 & 3   \\
RW Aur B &  K5         & 7.13$\pm$0.18  &  0.93$\pm$0.09  & ...   & 4395       & $-$0.40$\pm$0.10    & 1.56$\pm$0.24  \
  & 140 & 3   \\
\enddata
\tablecomments{A$_{v}$ = 1.11 A$_{\rm V}$. \\
1. Ingleby et al. (2013).  Their value A$_{V}$ = 0.5 (A$_{v}$ = 0.55) is based on 
   A$_{J}$ = 0.14 from  Furlan et al. (2011), which in turn is based on A$_{v}$ = 0.39 from White \& Ghez (2001). \\ 
2. Furlan et al. (2011) \\
3. White \& Ghez (2001). A$_{v}$ is from spectral type and $V-I_{C}$ color.
}
\end{deluxetable}

                                                                                                                                             
\begin{deluxetable}{lllcccccl}
\tabletypesize{\scriptsize}
\tablewidth{0pt}
\tablecaption{ X-ray Properties of RW Aur (Chandra ACIS-S)}
\tablehead{
         \colhead{Name} &
           \colhead{R.A.}               &
           \colhead{decl.}              &
           \colhead{Net Counts}         &
           \colhead{H.R.}               &
           \colhead{E$_{50}$}           &
           \colhead{P$_{const}$}             &
           \colhead{log L$_{x}$}             &
           \colhead{2MASS Identification(offset)}      \\
           \colhead{}   &
           \colhead{(J2000)}                 &
           \colhead{(J2000)} &
           \colhead{(cts)}                                          &
           \colhead{}                                          &
           \colhead{(keV)}                                          &
           \colhead{}                                          &
           \colhead{(ergs s$^{-1}$)}                                          &
           \colhead{(arcsec)}
                                  }
\startdata
RW Aur A         & 05 07 49.54 & $+$30 24 04.95 & 986 $\pm$ 33  & $-$0.72 & 1.00 & 0.96     & 29.44 & J050749.538$+$302405.07 (0.12) \\
RW Aur B         & 05 07 49.43 & $+$30 24 04.55 & 2909 $\pm$ 33 & $-$0.73 & 1.12 & $<$0.001 & 29.89 & ... \\
\enddata
\tablecomments{
The nominal pointing position for
the observation was (J2000.0) RA = 05$^h$ 07$^m$ 50.33$^s$,
decl. = $+$30$^{\circ}$ 23$'$ 59$''$.7.
X-ray data are from CCD7 (ACIS chip S3) using events in the 0.2 - 8 keV range inside a
circular source extraction region of radius 0$''$.9 centered on the X-ray peak of each star.
Events common to both sources (33 events) where the two circles overlap are excluded.
Tabulated quantities are: J2000.0 X-ray position (R.A., decl.), total source counts accumulated
in a 54,482 s exposure,  hardness ratio H.R. = (H$-$S)/(H$+$S) where H = counts(2-8 keV)
and S = counts(0.2 - 2 keV), median photon  energy (E$_{50}$),
probability of constant count-rate determined by the Gregory-Loredo
algorithm (P$_{const}$); unabsorbed X-ray luminosity (0.3 - 8 keV at d = 140 pc; L$_{x}$ is model-dependent, see  Table 3),
and 2MASS near-IR  counterpart identification.
The offset (in parenthesis) is given in arc seconds between the X-ray and 2MASS counterpart position.
The X-ray offset of B from A is 1.$''$48 along P.A. = 254.3$^{\circ}$.}

\end{deluxetable}

\newpage

\begin{deluxetable}{llllll}
\tabletypesize{\scriptsize}
\tablewidth{0pc}
\tablecaption{{\em Chandra} Spectral Fits of RW Aur A and B
   \label{tbl-1}}
\tablehead{
\colhead{Parameter}      &
\colhead{ }              &
\colhead{  }
}
\startdata
Object                                             & RW Aur A              & RW Aur A                 & RW Aur A                  & RW Aur B                  & RW Aur B     \nl
Model\tablenotemark{a}                             & 2T\tablenotemark{a}   & 2T\tablenotemark{a}      & 1T $+$ PL\tablenotemark{b}& 2T\tablenotemark{a}       & 2T\tablenotemark{a}  \nl
N$_{\rm H}$ (10$^{20}$ cm$^{-2}$)                  & 0.89 [0.00 - 6.50]    & \{6.5\}\tablenotemark{c} & \{6.5\}\tablenotemark{c}  & 4.3 [2.3 - 6.5]           & \{22.4\}\tablenotemark{d} \nl
kT$_{1}$ (keV)                                     & 0.73 [0.61 - 0.81]    & 0.65 [0.58 - 0.75]       & 0.73 [0.64 - 0.79]        & 0.98 [0.82 - 1.01]        & 0.39 [0.35 - 0.42] \nl
kT$_{2}$ (keV)                                     & $>$20.\tablenotemark{e}   & \{20.\}\tablenotemark{f} & ...                       & 3.15 [2.58 - 4.06]        & 2.39 [2.09 - 2.74] \nl
$\Gamma_{ph}$                                      & ...                   & ...                      & $-$0.04 [$-$0.76 - 0.59]  & ...                       & ... \nl
norm$_{1}$ (10$^{-5}$)\tablenotemark{g}            & 2.71 [1.89 - 3.86]    & 3.30 [2.46 - 4.15]       & 5.03 [4.14 - 5.84]        & 6.80 [4.35 - 9.85]        & 14.0 [11.4 - 16.8] \nl
norm$_{2}$  (10$^{-5}$)\tablenotemark{g}           & 3.56 [2.61 - 4.25]    & 2.86 [2.29 - 3.42]       & 0.10 [0.03 - 0.26]        & 13.3 [11.1 - 15.3]        & 18.4 [16.8 - 19.9] \nl
Ne                                                 & 1.39 [0.10 - 3.00]    & 1.53 [0.72 - 2.6]        & 0.76 [0.00 - 1.59]        & 1.69 [0.09 - 3.30]        & 2.16 [1.73 - 2.69] \nl
Fe                                                 & 0.41 [0.29 - 0.60]    & 0.39 [0.29 - 0.57]       & 0.27 [0.21 - 0.35]        & 0.36 [0.24 - 0.50]        & 0.22 [0.14 - 0.33] \nl
$\chi^2$/dof                                       & 78.5/62               & 83.3/64                  & 68.9/63                   & 116.1/124                 & 137.4/125          \nl
$\chi^2_{red}$                                     & 1.27                  & 1.30                     & 1.09                      & 0.94                      & 1.10 \nl
F$_{\rm X}$ (10$^{-13}$ ergs cm$^{-2}$ s$^{-1}$)   & 1.03 (1.06)           & 0.97 (1.17)   & 1.19 (1.39)               & 2.93 (3.31)\tablenotemark{h} & 2.75 (5.16)  \nl
F$_{\rm X,1}$ (10$^{-13}$ ergs cm$^{-2}$ s$^{-1}$) & 0.53 (0.55)           & 0.51 (0.67)   & 0.64 (0.83)               & 1.08 (1.27)               & 0.96 (2.55) \nl
F$_{\rm X,2}$ (10$^{-13}$ ergs cm$^{-2}$ s$^{-1}$) & 0.50 (0.51)           & 0.46 (0.50)   & 0.55 (0.56)               & 1.85 (2.04)               & 1.79 (2.61) \nl
log L$_{\rm X}$ (ergs s$^{-1}$)                    & 29.39                 & 29.44         & 29.51                     & 29.89                     & 30.08   \nl
log [L$_{\rm X}$/L$_{*}$]                          & $-$3.89               & $-$3.84       & $-$3.77                   & $-$3.29                   & $-$3.10 \nl
\enddata

\tablecomments{
Based on  XSPEC (vers. 12.7.1) fits of the background-subtracted ACIS-S spectra binned
to a minimum of 10 counts per bin using 54,482 sec of  exposure time. The spectra were
modeled with an absorbed  two-temperature (2T) $vapec$ optically thin plasma
model or an absorbed one-temperature (1T) $vapec$ plasma $+$ a power-law (PL) component.
The Ne and Fe abundances were allowed to vary during the fit but their abundances in the cool and hot
thermal plasma components were constrained  to be equal.
The tabulated parameters
are absorption column density (N$_{\rm H}$), plasma energy (kT),
photon power-law index ($\Gamma_{ph}$),
XSPEC component normalization (norm), neon (Ne) and iron (Fe) abundances.
Abundances are referenced to
the solar values of Anders \& Grevesse (1989).
Square brackets enclose 90\% confidence intervals.
Quantities enclosed in curly braces were held fixed during fitting.
The total X-ray flux (F$_{\rm X}$) and fluxes associated with each model component
(F$_{\rm X,i}$)  are the absorbed values in the 0.3 - 8 keV range, followed in
parentheses by  unabsorbed values.
The total X-ray luminosity L$_{\rm X}$  is the  unabsorbed
value in the 0.3 - 8 keV range and  assumes a
distance of 140 pc. The adopted stellar luminosities are  L$_{*}$ = 0.5 L$_{\odot}$ for
RW Aur A (Ingleby et al. 2013) and L$_{*}$ = 0.4  L$_{\odot}$ for RW Aur B (White \& Ghez 2001).}
\tablenotetext{a}{XSPEC model of form $wabs$$\cdot$($vapec$ $+$ $vapec$) .}
\tablenotetext{b}{XSPEC model of form $wabs$$\cdot$($vapec$ $+$ $pow$) .}
\tablenotetext{c}{Held fixed during fitting at upper 90\% confidence bound.}
\tablenotetext{d}{Held fixed during fitting at the value inferred from A$_{\rm V}$ = 1.4 mag
                  (A$_{v}$ = 1.56; White \& Ghez 2001) and the A$_{\rm V}$ to N$_{\rm H}$ conversion of Vuong et al. (2003).}
\tablenotetext{e}{Lower 90\% confidence bound. Hot-component temperature is poorly-constrained.}
\tablenotetext{f}{Value held fixed during fitting at lower 90\% confidence bound.}
\tablenotetext{g}{For thermal $vapec$ models, the norm is related to the volume emission measure
                  (EM = n$_{e}^{2}$V)  by
                  EM = 4$\pi$10$^{14}$d$_{cm}^2$$\times$norm, where d$_{cm}$ is the stellar
                  distance in cm. At d = 140 pc this becomes
                  EM = 2.35$\times$10$^{56}$ $\times$ norm (cm$^{-3}$). }
\tablenotetext{h}{The flux is variable. The 0.3 - 8 keV fluxes during the first and second halves
                  of the observation in units of 10$^{-13}$ ergs cm$^{-2}$ s$^{-1}$
                  were F$_{\rm X}$  = 2.33 (2.65) and F$_{\rm X}$ = 3.48 (3.90)
                  where the value in parentheses is unabsorbed. The corresponding
                  X-ray luminosities are log L$_{x}$ = 29.79 and 29.96 (ergs s$^{-1}$).}
\end{deluxetable}


\clearpage


\begin{figure}
\figurenum{1}
\epsscale{1.0}
\includegraphics*[width=7.0cm,height=4.83cm,angle=0]{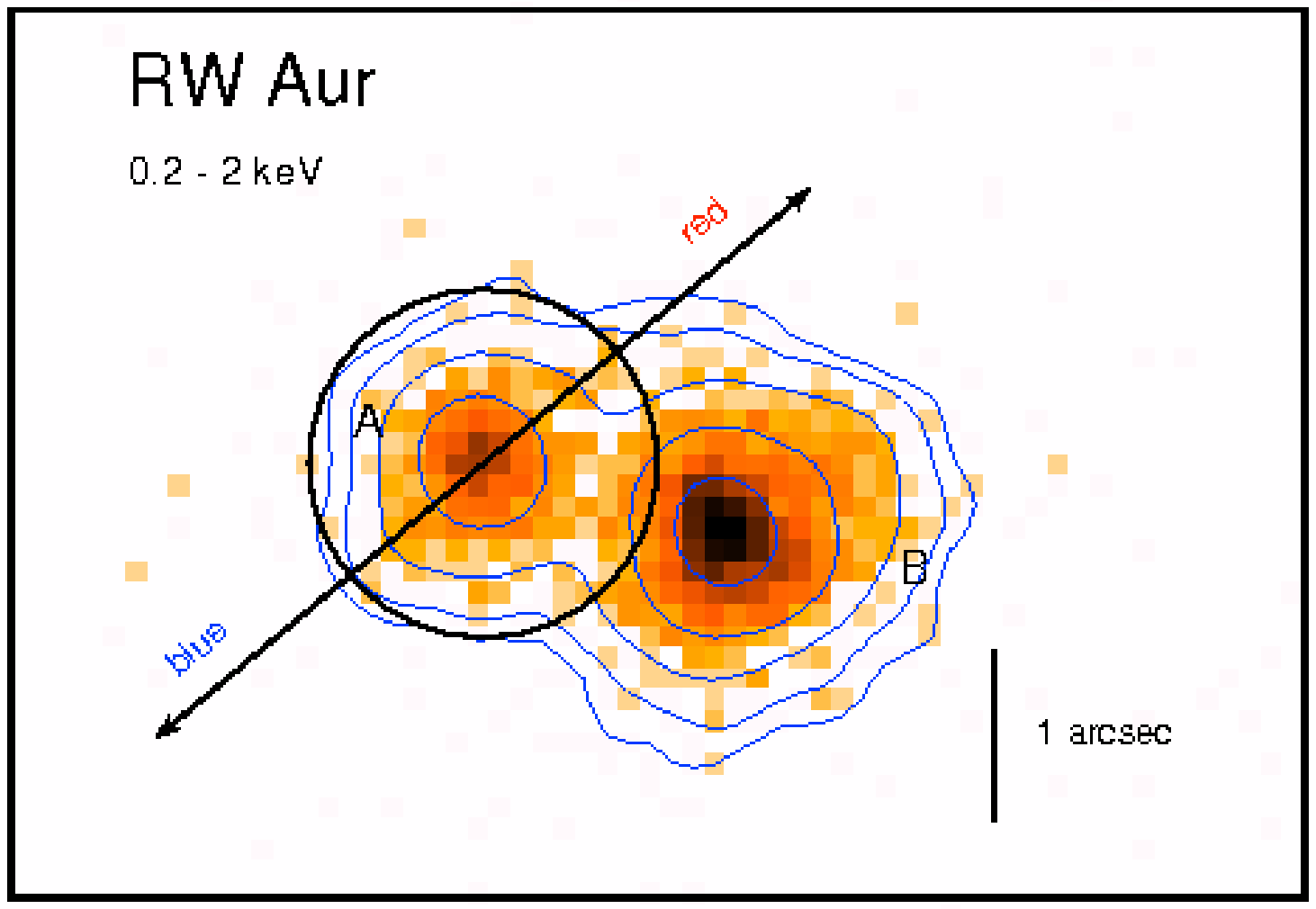}
\includegraphics*[width=7.0cm,height=4.83cm,angle=0]{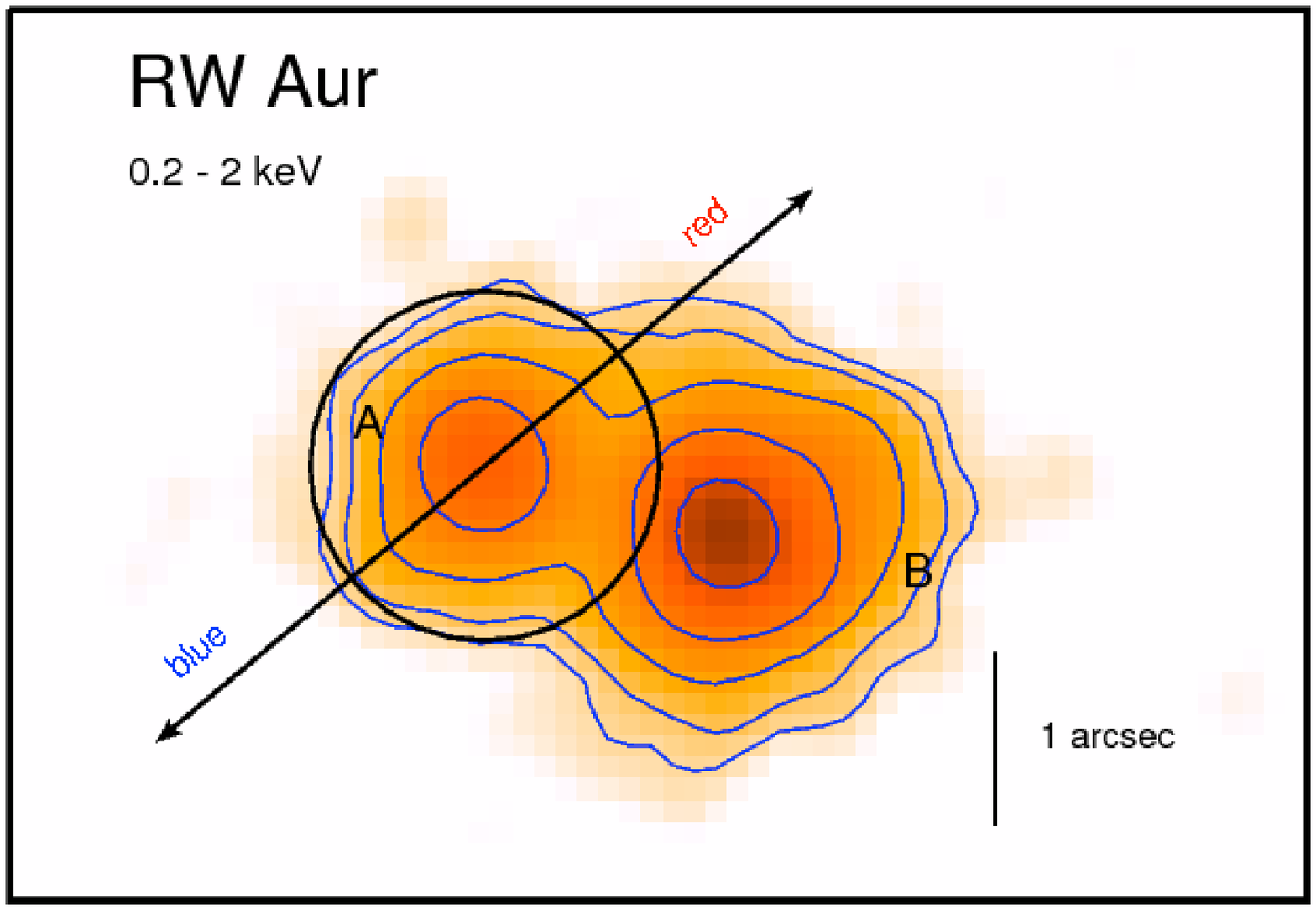} \\
\caption{Soft-band (0.2 - 2 keV)  ACIS-S image of RW Aur generated
using 0.$''$125 subpixels and log intensity scale. No deconvolution. The solid black circle around
component A has radius = 1$''$. In the 0.2 - 2 keV range, source A has 876 events and
B has 2536 events based on circular extraction regions of radius = 0.9$''$ centered on each
source. Of these, 24 events lie in the region where the two extraction circles overlap.
Arrows show the bipolar optical jet direction along P.A. = 130$^{\circ}$/310$^{\circ}$.
Slight extension of the outer contours is visible along the blueshifted jet axis.
{\em Left}:~Unsmoothed.~{\em Right}:~Gaussian-smoothed using a 3-subpixel kernel.
}
\end{figure}


\clearpage

\begin{figure}
\figurenum{2}
\epsscale{1.0}
\includegraphics*[width=7.0cm,height=4.83cm,angle=0]{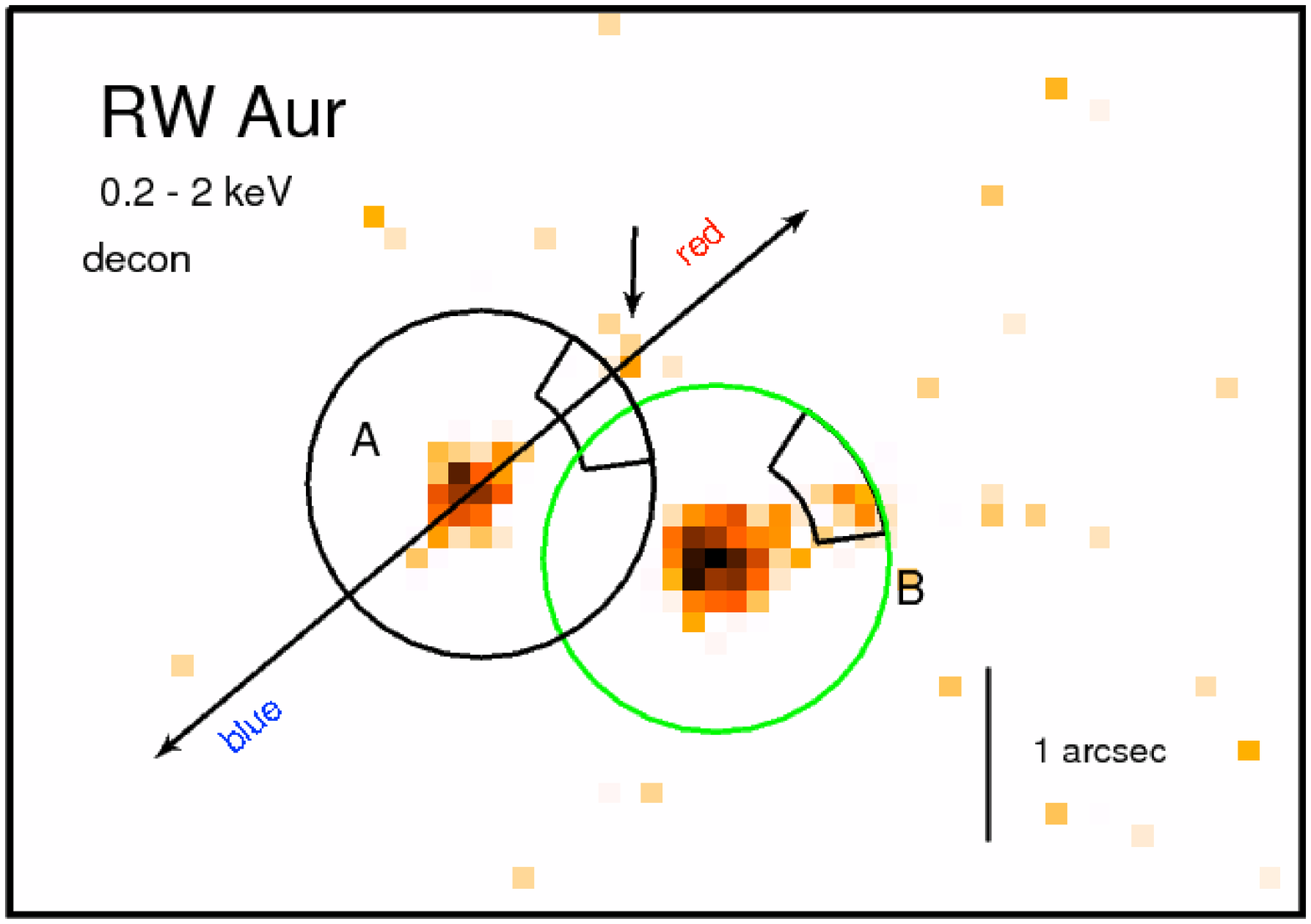}  
\includegraphics*[width=7.0cm,height=4.83cm,angle=0]{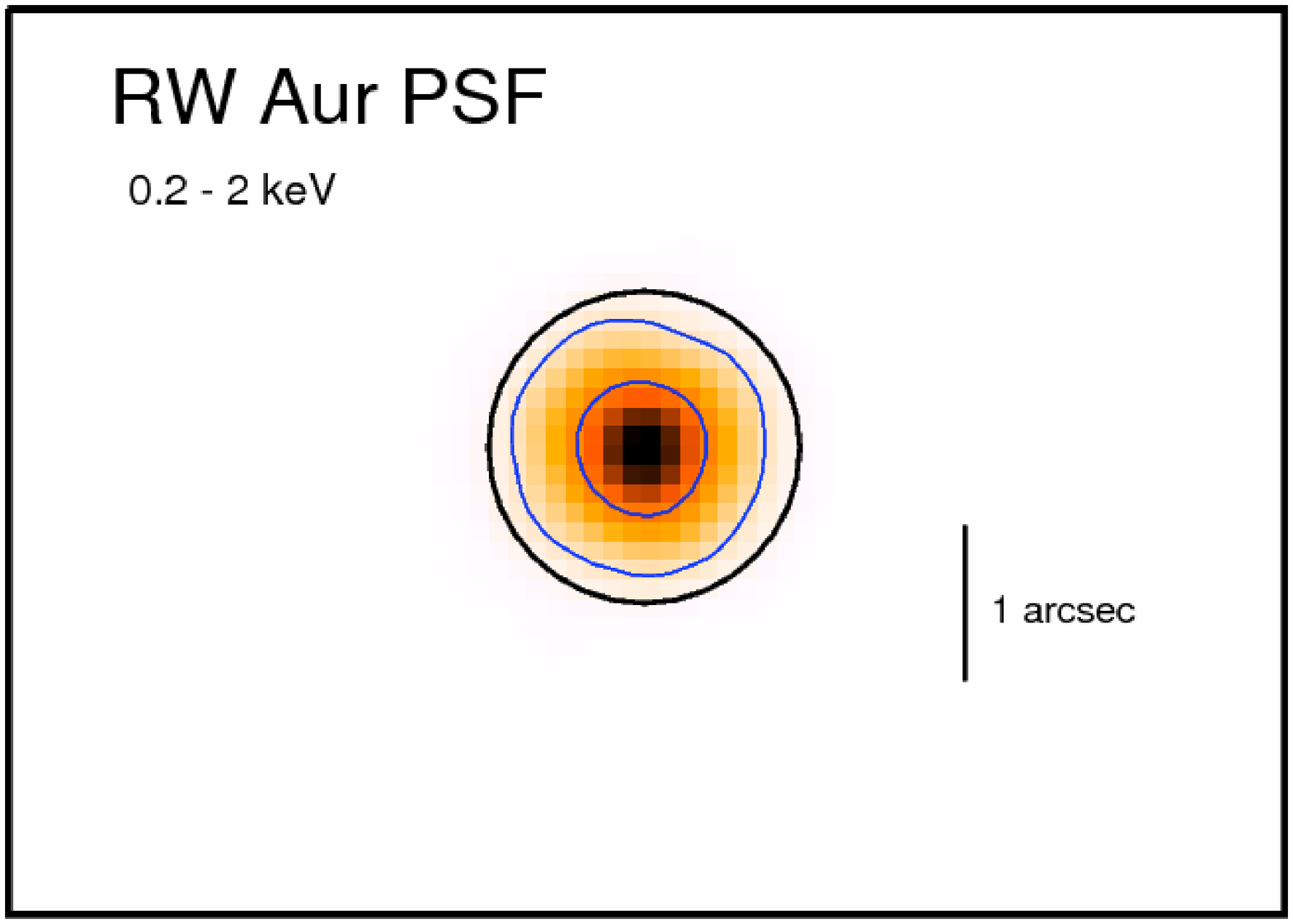} \\
\hspace{0.14in}
\includegraphics*[width=6.7cm,height=5.36cm,angle=0]{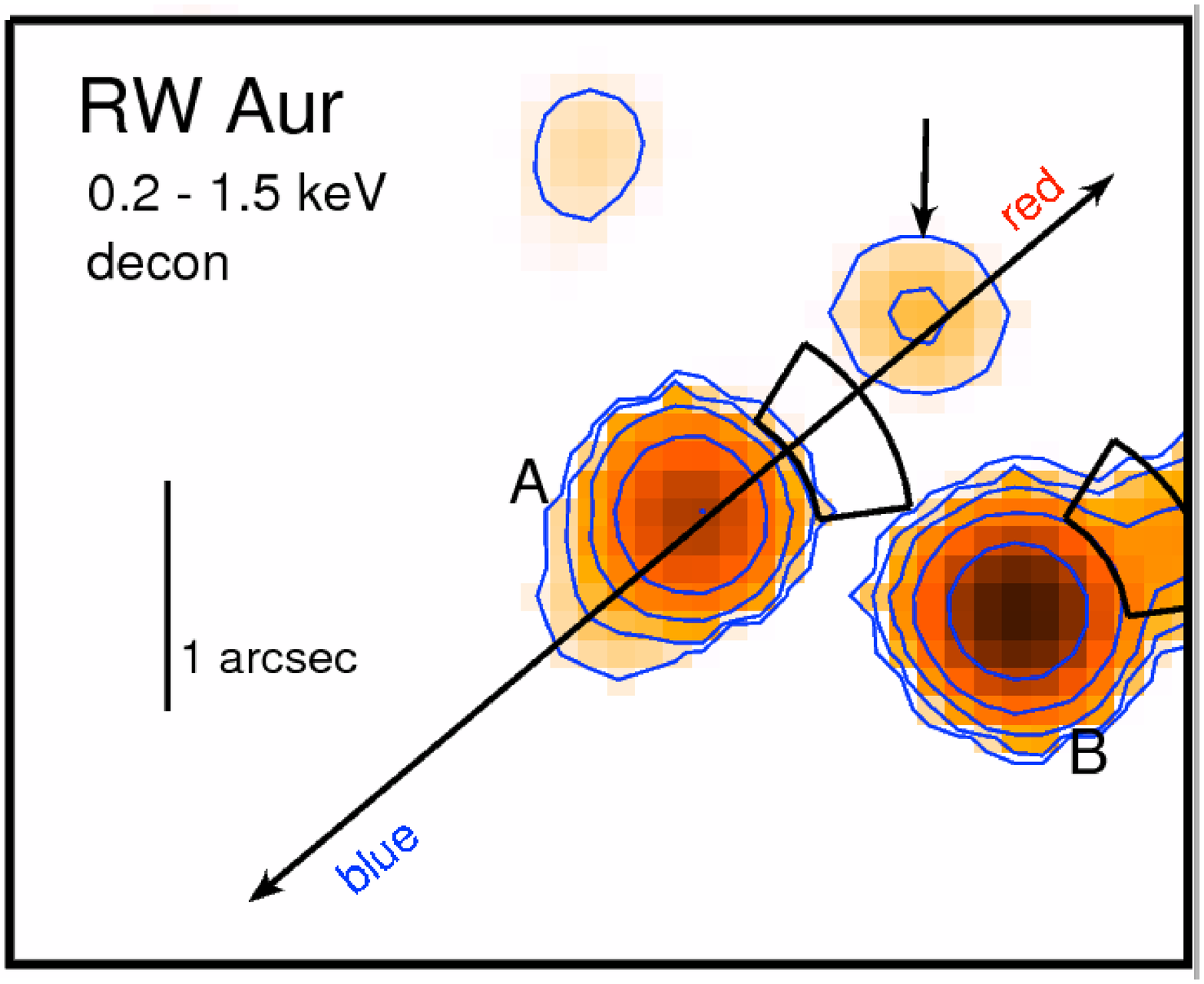}
\hspace{0.14in}
\includegraphics*[width=6.7cm,height=5.36cm,angle=0]{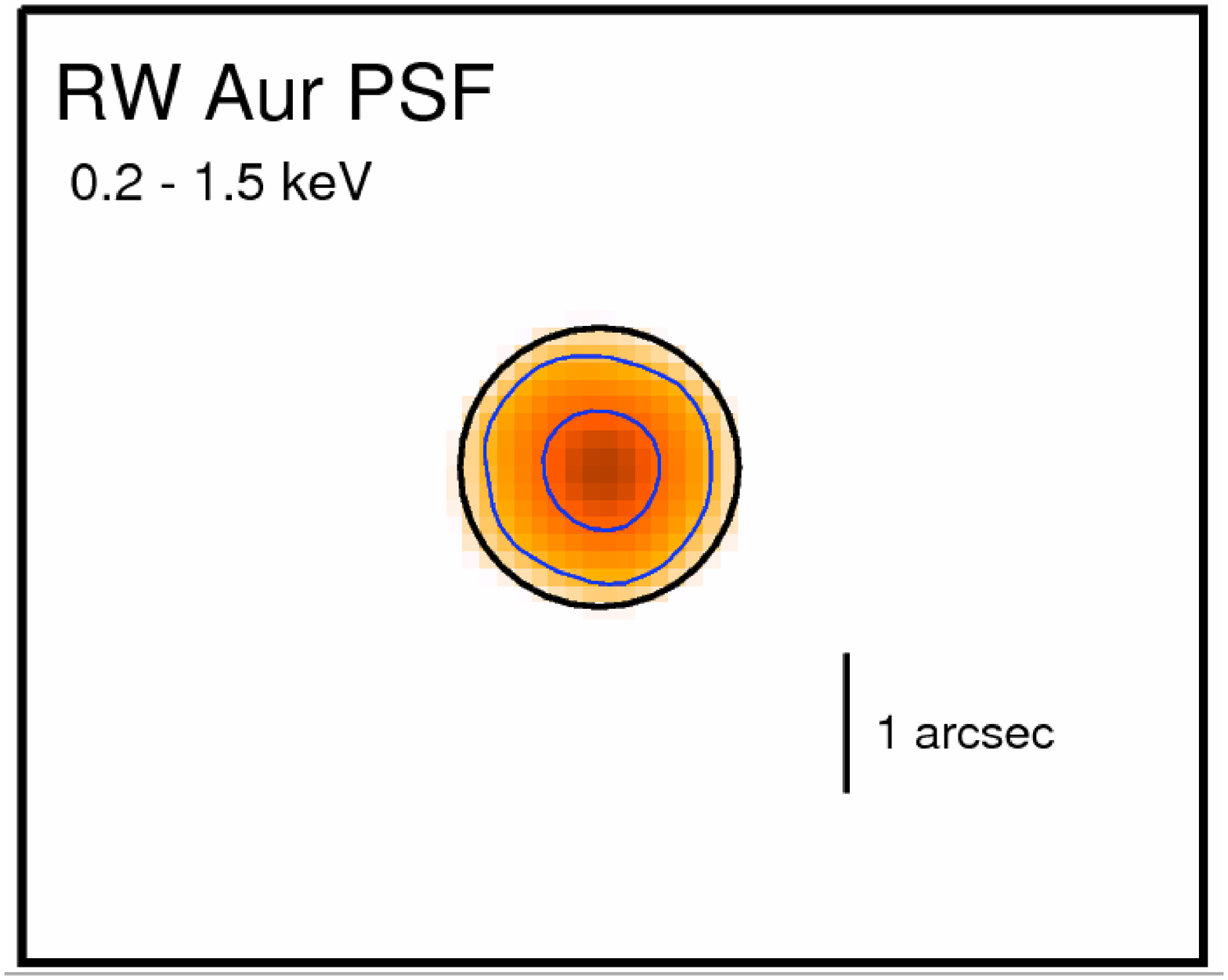}

\caption{{\em Top Left}:~Deconvolved unsmoothed soft-band (0.2 - 2 keV)  ACIS-S image of RW Aur
with 0.$''$125 subpixels  generated using the Lucy-Richardson method as implemented in 
the CIAO task $arestore$ (100 iterations). The deconvolution was centered on the 
brighter source B using the  PSF in the right panel. Log intensity scale. 
The sector regions (PA = 283$^{\circ}$ $\pm$ 25$^{\circ}$; radius 0.6$''$ $\leq r \leq$ 1$''$)
may contain  artificial structure due to a known PSF asymmetry. There is $<$1 count in sector A
and $\approx$41 counts in sector B. 
The arrow marks a faint emission peak (11$\pm$2 counts)  on the redshifted jet-axis of 
RW Aur A at an offset of 1.$''$2 $\pm$ 0.$''$2 from the star.
{\em Top Right}:~
 Chandra PSF image (0.2 - 2 keV; Gaussian-smoothed; log intensity scale), generated using RW Aur B
  spectrum as input (Fig. 4). Inner contour is half-maximum, outer contour is 10\% maximum.
  Outer circle has radius 1$''$. 
{\em Bottom Left}:~Same as top-left panel except using a more restrictive  0.2 - 1.5 keV energy cut and
              Gaussian-smoothed with a 3-subpixel kernel. RW Aur A contours are at levels of
              (0.3, 1.5, 6, 25)\% of maximum. Note slight outward extension along
              blueshifted jet axis.~
{\em Bottom Right}:~Same as top-right panel except 0.2 - 1.5 keV.
}
\end{figure}

\clearpage


\begin{figure}
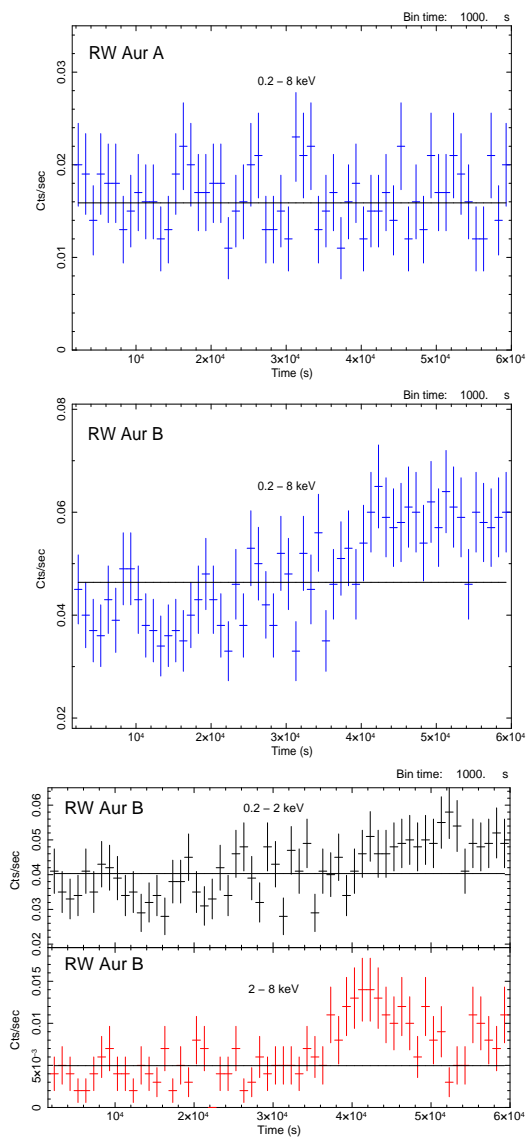

\figurenum{3}
\epsscale{1.0}
\includegraphics*[width=5.0cm,height=7.0cm,angle=-90]{f3t.eps} \\
\includegraphics*[width=5.0cm,height=7.0cm,angle=-90]{f3m.eps} \\
\includegraphics*[width=5.0cm,height=7.0cm,angle=-90]{f3b.eps} \\
\caption{Chandra ACIS-S light curves of RW Aur binned at 1000 s intervals.
The light curves were extracted from circular regions of radii 0.$''$9
centered on each source but events within the region 
between the stars where the circles overlap were excluded.
Solid horizontal lines show mean count rates.
{\em Top}:~
RW Aur A (0.2 - 8 keV). No significant variability is detected.~
{\em Middle }:~RW Aur B (0.2 - 8 keV); variability is clearly present in the 
  second half of the observation.~
{\em Bottom }:~RW Aur B soft (0.2 - 2 keV) and hard (2 - 8 keV).
}
\end{figure}

\clearpage


\begin{figure}
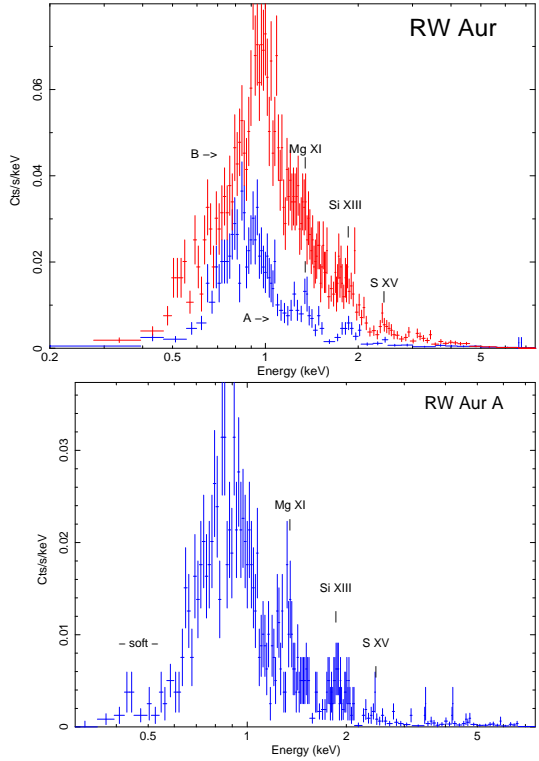

\figurenum{4}
\epsscale{1.0}
\includegraphics*[width=5.0cm,height=7.0cm,angle=-90]{f4t.eps} \\
\includegraphics*[width=5.0cm,height=7.0cm,angle=-90]{f4b.eps} \\
\caption{{\em Top}:~Chandra ACIS-S spectra  of RW Aur A and B binned to a minimum of 
 10 counts per bin. The spectra were extracted from circular regions of radii 0.$''$9 centered 
 on each source but events within the region between the stars where the circles overlap were excluded.
 Possible blended emission lines  are identified. 
{\em Bottom}:~Spectrum of RW Aur A, lightly-binned to a minimum of 2 counts per bin to
show faint emission (22 counts) below 0.5 keV. The spectrum was extracted from a rectangular slit region of
length 10$''$ and full-width 0.8$''$ oriented along the optical jet axes at 
P.A. = 130$^{\circ}$/310$^{\circ}$  and centered on the star. 
}
\end{figure}

\clearpage

\begin{figure}
\figurenum{5}
\epsscale{1.0}
\includegraphics*[width=5.0cm,height=7.0cm,angle=-90]{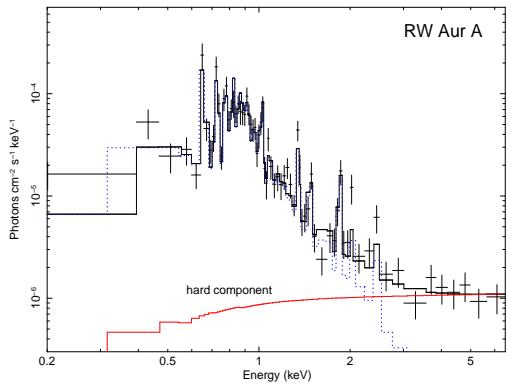} 
\caption{Unfolded spectrum of RW Aur A showing a fit (solid black line)
         with a 1T $apec$ $+$ power-law (PL)
         model (Table 3). The thermal $apec$ model (dashed blue line) reproduces the emission below
         $\approx$2.5 keV and the PL model (solid red line) accounts for the hard emission at higher energies.
         The hard component can also be reproduced using a very hot thermal plasma model (2T model in Table 3).
         The spectrum contains 986 counts (0.2 - 8 keV) of which 136 counts are in the hard 2 - 8 keV range.}
\end{figure}

\clearpage

\end{document}